\documentclass[11pt,twoside]{article}
\usepackage[pdftex]{graphicx}
\usepackage{amsmath}
\usepackage{amssymb}
\usepackage{cite}
\begin{document}
\newcommand{\pst}{\hspace*{1.5em}}

\newcommand{\rigmark}{\em Journal of Russian Laser Research}
\newcommand{\lemark}{\em Volume 30, Number 5, 2009}

\newcommand{\be}{\begin{equation}}
\newcommand{\ee}{\end{equation}}
\newcommand{\bm}{\boldmath}
\newcommand{\ds}{\displaystyle}
\newcommand{\bea}{\begin{eqnarray}}
\newcommand{\eea}{\end{eqnarray}}
\newcommand{\ba}{\begin{array}}
\newcommand{\ea}{\end{array}}
\newcommand{\arcsinh}{\mathop{\rm arcsinh}\nolimits}
\newcommand{\arctanh}{\mathop{\rm arctanh}\nolimits}
\newcommand{\bc}{\begin{center}}
\newcommand{\ec}{\end{center}}

\thispagestyle{plain}

\label{sh}


\begin{center} {\Large \bf
\begin{tabular}{c}
 THE REPLICA METHOD AND ENTROPY FOR MIXTURE OF
 \\[-1mm]
 TWO-MODE EVEN AND ODD SCHR\"{O}DINGER CAT STATES
\end{tabular}
 } \end{center}

\bigskip

\bigskip

\begin{center} {\bf
Ivan V. Dudinetc$^{1*}$ and Vladimir I. Man'ko $^{1,2}$
}\end{center}

\begin{center}
{\it
$^1$Moscow Institute of Physics and Technology (State University)\\
Institutskii per. 9, Dolgoprudnii, Moscow Region 141700, Russia

\smallskip

$^2$P.N. Lebedev Physical Institute, Russian Academy of Sciences\\
Leninskii Prospect 53, Moscow 119991, Russia
}
\smallskip

$^*$Corresponding author e-mail: dudinecivanvas@yandex.ru \\
\end{center}

\begin{abstract}\noindent
The replica method for calculating the von Neumann entropy is reviewed. Explicit expression for the entropy of the mixed coherent states $|\alpha\rangle$ and $|\beta\rangle$ is obtained using this method. The purity inequality for bipartite system for separable states is studied on example of even and odd Schr\"{o}dinger cat state.
\end{abstract}

\medskip


\section{Introduction}
\pst
The states of systems in quantum mechanics are associated either with wave functions~\cite{1} or density matrices~\cite{2,3,4}. The superposition of two wave functions provides the functions which correspond to other quantum states. Such superposition states were discussed in~\cite{5} and they are called Schr\"{o}dinger cat states. The quantum states providing the superposition can be chosen as, e.g. coherent states $|\alpha\rangle$~\cite{6,7}. Superposition of two coherent states $|\alpha\rangle$ and $|\beta\rangle$ was considered in~\cite{8}. Since the coherent states, being pure quantum states, are interpreted as closest to "classical states", the superposition of the coherent states are usually discussed as cat states. The example of such superposition is the even and odd superposition of the coherent states $|\alpha\rangle$ and $|-\alpha\rangle$. Even and odd coherent states for one-mode systems were introduced in~\cite{9}. They are studied in~\cite{10,11}. The multimode generalization of Schr\"{o}dinger cat states was considered in~\cite{12,13}. The properties of even and odd coherent states were discussed in detail in the literature~\cite{14,15,16,17,18,19}. Entangled coherent states were reviewed in Ref.~\cite{20}. The even and odd Schr\"{o}dinger cat states were discussed also in connection with using the states for the aim of quantum computations as analogs of qubit states~\cite{21}. In view of the possible application of the states in quantum information one needs to study such properties of the states as entropic and other quantum inequalities and obtain the corresponding relations in explicit form. 
The aim of this paper is to use the replica method~\cite{22} to find the explicit form of the von Neumann entropy for mixture of even and odd coherent states. We will also consider purity inequality for bipartite system~\cite{23} for mixture of two-mode coherent states.

The paper is organized as follows.

In Sec.~2 the replica method for calculating von Neumann entropy for mixture of one-mode coherent states is used. Two-mode Schr\"{o}dinger cat states are studied in Sec.~3. 
In Sec.~4 purity inequality for mixture of two-mode even and odd Schr\"{o}dinger cat states is considered. Finally, some brief conclusions are given in Sec.~5.

\section{The replica method for calculating the von Neumann entropy}
\pst
In this section we calculate the von Neumann entropy for one-mode mixture of coherent states.

For a quantum-mechanical system described by a density operator $\widehat{\rho}$, the von Neumann entropy is defined as
\be\label{Neum}
 S(\widehat{\rho})=-\mbox{Tr}(\widehat{\rho}\,\mbox{ln}\,\widehat{\rho}).
\ee
To compute the von Neumann entropy it is convenient  to diagonalize $\widehat{\rho}$,
then the von Neumann entropy is given by the expression 
\be\label{Neum2}
 S(\widehat{\rho})=-\sum_{k}\lambda _{k}\,\mbox{ln}\,\lambda _{k},
\ee
where $\lambda _{k}$ are the eigenvalues of density operator $\widehat{\rho}$. 
But in some cases to calculate the von Neumann entropy is difficult task because we need finding the eigenvalues of the matrix, which is the infinite-dimensional matrix corresponding to the operator $\widehat{\rho}$.
Nevertheless, we can circumvent this problem, using replica method.
We can rewrite the von Neumann entropy as~\cite{22}
\be\label{repl}
 S(\widehat{\rho})=-\lim_{n \to 1}\frac{\partial }{\partial n}\mbox{Tr}\,\widehat{\rho}^n.
\ee
So, we can avoid the need to diagonalise the density operator $\widehat{\rho}$.
The problem reduces to calculation of $\mbox{Tr}\widehat{\rho}^n$ as function of $n$.
In the work~\cite{22} the entropy of the thermal state, entropy of the state generated by amplifying a Fock state and entropy produced by amplifying a superposition of the vacuum and a Fock state were considered. In our work we apply the replica method to other interesting states.

Let us consider a density operator in the form
\be \label{rho}
 \widehat{\rho}=a\,|\alpha\rangle\langle\alpha|+c\,|\alpha\rangle\langle\beta|+c^*\,|\beta\rangle\langle\alpha|+ b\,|\beta\rangle\langle\beta|, 
\ee
where $a$ and $b$ are non-negative numbers, $c$ is a complex number.
The states $|\alpha\rangle$ and $|\beta\rangle$ are the eigenstates of the photon annihilation operators, i.e., $\widehat{a}\,|\alpha\rangle = \alpha\,|\alpha\rangle$, $\widehat{a}\,|\beta\rangle = \beta\,|\beta\rangle$.
In order to have the trace of $\widehat{\rho}$ equal to unity, the numbers $a$,\,$b$ and $c$ have to satisfy condition 
$a+c\,\langle\beta|\alpha\rangle+c^*\,\langle\alpha|\beta\rangle+b = 1$.
The scalar product $\langle\beta|\alpha\rangle$ of the coherent states $|\alpha\rangle$ and $|\beta\rangle$ has the value
\be \label{scal}
 \langle\beta|\alpha\rangle=e^{-\frac{1}{2}|\alpha|^2-\frac{1}{2}|\beta|^2+\beta^*\alpha},
\ee
which shows that two coherent states are not orthogonal.
The von Neumann entropy of the state~(\ref{rho}) can be easily calculated using replica method.
In order to apply replica method, we calculate the following trace $\mbox{Tr}(\widehat{\rho}^n)$. It is obvious that density operator 
$\widehat{\rho}$ to the power $n$ ($n$ is an arbitrary positive integer number) contains only four combinations of coherent state projectors $|\alpha\rangle$ and $|\beta\rangle$
\be \label{repl:n}
 \widehat{\rho}^n=C^{(n)}_1|\alpha\rangle\langle\alpha|+C^{(n)}_2|\alpha\rangle\langle\beta|+C^{(n)}_3|\beta\rangle\langle\alpha|+C^{(n)}_4|\beta\rangle\langle\beta|,
\ee
where $C^{(n)}_i$, $i=1$, $\ldots\,$,~4 are complex numbers, which can be easily found recursively. Initial condition~(\ref{rho}) yields
$\left (C^{(1)}_1,\,C^{(1)}_2,\,C^{(1)}_3,\,C^{(1)}_4 \right )=
\left (a,\,c,\,c^*,\,b \right )$.
It is not hard to prove the recurrence relation for coefficients $C^{(n)}_i$ from~(\ref{repl:n})
\be
\begin{pmatrix}\label{C1}C^{(n)}_1
\\ C^{(n)}_2
\\ C^{(n)}_3
\\ C^{(n)}_4

\end{pmatrix}=\begin{pmatrix}
a+c^*\,\langle\alpha|\beta\rangle & a\,\langle\beta|\alpha\rangle+c^* & 0 & 0\\ 
c+b\,\langle\alpha|\beta\rangle &  c\,\langle\beta|\alpha\rangle+b& 0 & 0\\ 
0 & 0 & a+c^*\,\langle\alpha|\beta\rangle & a\,\langle\beta|\alpha\rangle+c^*\\ 
0 & 0 & c+b\,\langle\alpha|\beta\rangle & c\,\langle\beta|\alpha\rangle+b
\end{pmatrix}
\begin{pmatrix}C^{(n-1)}_1
\\ C^{(n-1)}_2
\\ C^{(n-1)}_3
\\ C^{(n-1)}_4

\end{pmatrix}.
\ee
Formula~(\ref{C1}) leads to the following expression for $C^{(n)}_i$
\be
\begin{pmatrix}\label{C2}C^{(n)}_1
\\ C^{(n)}_2
\\ C^{(n)}_3
\\ C^{(n)}_4

\end{pmatrix}=\begin{pmatrix}
\begin{pmatrix}
 a+c^*\,\langle\alpha|\beta\rangle & a\,\langle\beta|\alpha\rangle+c^* \\ 
 c+b\,\langle\alpha|\beta\rangle &  c\,\langle\beta|\alpha\rangle+b 
\end{pmatrix}^{n-1} & \begin{matrix}
0 &&&&&& 0 \\ 
0 &&&&&& 0
\end{matrix}\\ 
\begin{matrix}
0 &&&&&& 0 \\ 
0 &&&&&& 0
\end{matrix} & \begin{pmatrix}
 a+c^*\,\langle\alpha|\beta\rangle & a\,\langle\beta|\alpha\rangle+c^* \\ 
 c+b\,\langle\alpha|\beta\rangle &  c\,\langle\beta|\alpha\rangle+b 
\end{pmatrix} ^{n-1}
\end{pmatrix}
\begin{pmatrix}a
\\ c
\\ c^*
\\ b
\end{pmatrix}.
\ee
Combining~(\ref{repl:n}) and~(\ref{C2}), we get the formula for $\mbox{Tr}(\widehat{\rho}^n)$ of density operator~(\ref{rho})
\be\label{trace}
  \mbox{Tr}\,\widehat{\rho}^n=C^{(n)}_1+C^{(n)}_4+\langle\beta|\alpha\rangle\,C^{(n)}_2+\langle\alpha|\beta\rangle\,C^{(n)}_3
= \mbox{Tr}\begin{pmatrix}
 a+c^*\,\langle\alpha|\beta\rangle & a\,\langle\beta|\alpha\rangle+c^*\, \\ 
c+b\,\langle\alpha|\beta\rangle &  c\,\langle\beta|\alpha\rangle+b \\ 
\end{pmatrix}^{n}=\lambda_1^n+\lambda_2^n,
\ee
where the eigenvalues $\lambda_{1}$ and $\lambda_{2}$ are solutions of quadric equation $\lambda^2-\lambda + D(\alpha,\beta) = 0$, where $D(\alpha,\beta) = \left ( ab-|c|^2 \right )\left ( 1-e^{-|\alpha-\beta|^2} \right )$, i.e.,
\begin{align}\label{quadr}
&\lambda_{1}=\frac{1 + \sqrt{1-4 D(\alpha,\beta)}}{2},&
 \lambda_{2}=\frac{1 - \sqrt{1-4 D(\alpha,\beta)}}{2}.
\end{align}
Then, we readily find from~(\ref{repl}) the von Neumann entropy of the state~(\ref{rho})
\be\label{entr:lam}
S(\widehat{\rho})=-\lambda_1\,\mbox{ln}\,\lambda_1-\lambda_2\,\mbox{ln}\,\lambda_2.
\ee
In terms of the parameters determining the density operator~(\ref{rho}) the entropy $S(\widehat{\rho})$ reads
\begin{eqnarray}\label{entrexpl}
S(\widehat{\rho})=-\frac{1 + \sqrt{1-4 \left ( ab-|c|^2 \right )\left ( 1-e^{-|\alpha-\beta|^2} \right )}}{2}\,\mbox{ln}\,\left (\frac{1 + \sqrt{1-4 \left ( ab-|c|^2 \right )\left ( 1-e^{-|\alpha-\beta|^2} \right )}}{2}  \right )- \nonumber \\
\frac{1 - \sqrt{1-4 \left ( ab-|c|^2 \right )\left ( 1-e^{-|\alpha-\beta|^2} \right )}}{2}\,\mbox{ln}\,\left (\frac{1 - \sqrt{1-4 \left ( ab-|c|^2 \right )\left ( 1-e^{-|\alpha-\beta|^2} \right )}}{2}  \right ).
\end{eqnarray}
We can see that for $\alpha = \beta$ or $|c|^2 = a\,b$ (in this cases the state~(\ref{rho}) is pure) solutions of quadric equation~(see Eq. \ref{quadr}) are $\lambda_1 = 1$ and $\lambda_2 = 0$, hence $S(\widehat{\rho}) = 0$. 

Thus, we obtained the explicit form of entropy of the state~(\ref{rho}) using the replica method. 

\section{Entropy of cat states for two-mode system}
\pst
Let us consider a bipartite system with density operator $\widehat{\rho}\,(1,2)$ depending on continuous variables of the first subsystem 1 and the second subsystem 2.
The density operators of two subsystems $\widehat{\rho}\,(1)$ and $\widehat{\rho}\,(2)$ are given as partial traces
\begin{align}\label{reduc}
&\widehat{\rho}\,(1) = \mbox{Tr}_2\,\widehat{\rho}\,(1,2),&
 \widehat{\rho}\,(2) = \mbox{Tr}_1\,\widehat{\rho}\,(1,2).
\end{align}
The operators $\widehat{\rho}\,(1)$ and $\widehat{\rho}\,(2)$ describe the state of subsystems 1 and 2, respectively.

Let us apply the replica method to the mixture of two-mode Schr\"{o}dinger cat states
\be\label{cats}
 \widehat{\rho}\,(1,2)=a\,|\overrightarrow{\alpha_+}\rangle\langle\overrightarrow{\alpha_+}|+b\,|\overrightarrow{\alpha_-}\rangle\langle\overrightarrow{\alpha_-}|
\ee
where $a$ and $b$ are real positive numbers with sum equal to unit, i.e., $a+b = 1$ and two-mode Schr\"{o}dinger cat states are defined as $|\overrightarrow{\alpha_{\pm }}\rangle = N_{\pm }\left ( |\alpha_1,\,\alpha_2\rangle \pm |-\alpha_1,\,-\alpha_2\rangle \right )$, the normalization constants read~\cite{12}
$N_{\pm} =2^{-1/2}\left ( 1 \pm e^{-2|\alpha_1|^2-2|\alpha_2|^2} \right )  ^{-1/2}$. Taking advantage of equality $\langle\overrightarrow{\alpha_-}|\overrightarrow{\alpha_+}\rangle = 0$ the von Neumann entropy of the bipartite system~(\ref{cats}) reads $S(\widehat{\rho}\,(1,2)) = -a\,\mbox{ln}\,a-b\,\mbox{ln}\,b$. The von Neumann entropies of the reduced density operators $S(\widehat{\rho}\,(1))$ and $S(\widehat{\rho}\,(2))$ have the form~(\ref{quadr},~\ref{entr:lam}), where $D(\alpha,\beta)$ must be substituted by $D(\alpha_1,\alpha_2) =
\left ( \left ( a\,N^2_{+}+b\,N^2_{-} \right )^2 - e^{-4|\alpha_2|^2}\left ( a\,N^2_{+}-b\,N^2_{-} \right )^2 \right )\left ( 1-e^{-4|\alpha_1|^2} \right )$ for $S(\widehat{\rho}\,(1))$ and by $D(\alpha_2,\alpha_1)$ for $S(\widehat{\rho}\,(2))$. The final expression for the entropy of the first subsystem $S(\widehat{\rho}\,(1))$ given in terms of parameters of~(\ref{cats}) reads
\begin{eqnarray}\label{entr1}
S(\widehat{\rho}\,(1))=-\frac{1 + \sqrt{1-4 D(\alpha_1,\alpha_2)}}{2}\,\mbox{ln}\,\left ( \frac{1 + \sqrt{1-4 D(\alpha_1,\alpha_2)}}{2} \right )-\frac{1 - \sqrt{1-4 D(\alpha_1,\alpha_2)}}{2} \nonumber \times\\
\,\mbox{ln}\,\left ( \frac{1 - \sqrt{1-4 D(\alpha_1,\alpha_2)}}{2} \right ),
\end{eqnarray}
where
\begin{eqnarray}\label{De}
D(\alpha_1,\alpha_2) =
\frac{1}{4}\left ( 1-e^{-4|\alpha_1|^2} \right )
\left ( a\,\left ( 1 + e^{-2|\alpha_1|^2-2|\alpha_2|^2} \right )  ^{-1}+b\,\left ( 1 - e^{-2|\alpha_1|^2-2|\alpha_2|^2} \right )  ^{-1} \right )^2 -\nonumber \\
\frac{1}{4}\,e^{-4|\alpha_2|^2}\left ( 1-e^{-4|\alpha_1|^2} \right )\left ( a\,\left ( 1 + e^{-2|\alpha_1|^2-2|\alpha_2|^2} \right )  ^{-1}-b\,\left ( 1 - e^{-2|\alpha_1|^2-2|\alpha_2|^2} \right )  ^{-1} \right )^2.
\end{eqnarray}

In Fig. 1 we have plotted the von Neumann entropy of the first subsystem $S(\widehat{\rho}\,(1))$ versus $|\alpha_1|$ for various ratio $|\alpha_1|/|\alpha_2|$. For small values of $|\alpha_i|$, i.e., $|\alpha_i|\ll 1$ , $i=1$,~2, the reduced density operator $\widehat{\rho}\,(1)$ of the system~(\ref{cats}) becomes 
\be
 \widehat{\rho}\,(1) =
\left (a+\frac{b|\alpha_2|^2}{|\alpha_1|^2+|\alpha_2|^2}  \right )|0\rangle\langle0|+\frac{b|\alpha_1|^2}{|\alpha_1|^2+|\alpha_2|^2}|1\rangle\langle1|, 
\ee
 ($|0\rangle$ and $|1\rangle$ are the eigenstates of the first mode photon number operator $\widehat{a}_1^+\widehat{a}_1$ with $\widehat{a}_1^+$ and $\widehat{a}_1$ being the photon creation and annihilation operators, respectively). Therefore, the von Neumann entropy of the first subsystem for small parameters $|\alpha_i|$ tends to
\be
 S(\widehat{\rho}\,(1))=-\left (a+\frac{b|\alpha_2|^2}{|\alpha_1|^2+|\alpha_2|^2}  \right )\mbox{ln}\left (a+\frac{b|\alpha_2|^2}{|\alpha_1|^2+|\alpha_2|^2}  \right )-\frac{b|\alpha_1|^2}{|\alpha_1|^2+|\alpha_2|^2}\,\mbox{ln}\left (\frac{b|\alpha_1|^2}{|\alpha_1|^2+|\alpha_2|^2}  \right )
\ee
 and for $|\alpha_i|\gg 1$, $i=1$,~2 becomes 
\be 
\widehat{\rho}\,(1) = \frac{1}{2}|\alpha_1\rangle\langle \alpha_1|+\frac{1}{2}|-\alpha_1\rangle\langle -\alpha_1|.
 \ee
Hence entropy $S(\widehat{\rho}\,(1))$ for large parameters $|\alpha_i|$ tends to $\mbox{ln}\,2$.
  
\begin{figure}[ht]
\bc \includegraphics[width=8.6cm]{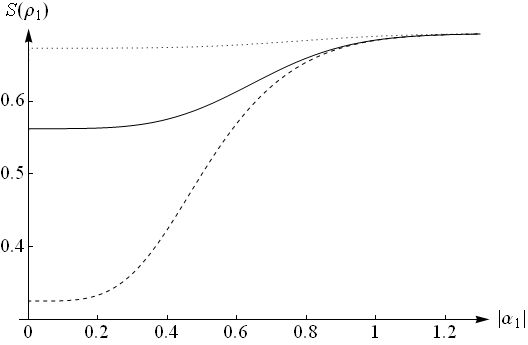}\ec
\caption{The von Neumann entropy of reduced density operator $\widehat{\rho}\,(1)$ versus $|\alpha_1|$ for $|\alpha_2| = |\alpha_1|$ (solid line), $|\alpha_2|=0.5\,|\alpha_1|$ (dotted line) and $|\alpha_2| = 2\,|\alpha_1|$ (dashed line); $a=b=0.5$.}
\end{figure}

\section{Purity inequality for bipartite system}
\pst
In this section we will consider the purity inequality for bipartite system for two - mode mixture of coherent states and mixture of coherent and a thermal state. 

Let us consider bipartite system with density operator $\widehat{\rho}\,(1,2)$. 
Let define purity parameters  of the system $\widehat{\rho}\,(1,2)$ and the first and second subsystems
\begin{align}\label{mu}
&\mu\,(1,2)=\mbox{Tr}\,\widehat{\rho}^2\,(1,2),
&\mu\,(1)=\mbox{Tr}_2\,\widehat{\rho}^2\,(1),
&&\mu\,(2)=\mbox{Tr}_1\,\widehat{\rho}^2\,(2),
\end{align}
where $\widehat{\rho}\,(1)$ and $\widehat{\rho}\,(2)$ are the density operators of first and second subsystems, respectively.  
For arbitrary finite density matrix corresponding to the operator $\widehat{\rho}\,(1,2)$ purity parameters $\mu\,(1,2)$, $\mu\,(1)$ and $\mu\,(2)$ satisfy inequality~\cite{23}
\be\label{ineq:mu}
1+\mu\,(1,2)\geq \mu\,(1)+\mu\,(2).
\ee

Let us consider the density operator of the separable cat state in the form
\be\label{two:mode}
\widehat{\rho}\,(1,2)=a\,|\alpha_1,\alpha_2\rangle\langle\alpha_1,\alpha_2|+ b\,|-\alpha_1,-\alpha_2\rangle\langle-\alpha_1,-\alpha_2|,
\ee
where real positive numbers $a$ and $b$ satisfy condition $a+b=1$ and  $|\alpha_1,\alpha_2\rangle = |\alpha_1\rangle|\alpha_2\rangle$. The
states $|\alpha_i\rangle$ are the eigenstates of the photon annihilation operators of $i$th mode, i.e.,
$\widehat{a}_i|\alpha_i\rangle = \alpha_i|\alpha_i\rangle$, $i=1$,~2 (coherent states of the first and the second modes, respectively). One can obtain explicit expressions for the purity parameters $\mu\,(1,2)$, $\mu\,(1)$ and $\mu\,(2)$.
Then the inequality~(\ref{ineq:mu}) for density operator~(\ref{two:mode}) reads
\be\label{ineq:prov}
 1+\mu\,(1,2)-\mu\,(1)-\mu\,(2)=2\,a\,b\,\left (1- e^{-4|\alpha_1|^2} \right )\left (1- e^{-4|             \alpha_2|^2} \right )\geq  0,
\ee
where we took into account that the density operator for the first and the second subsystem are given as partial traces $\widehat{\rho }\,(i) = a\,|\alpha_i\rangle\langle\alpha_i|+ b\,|-\alpha_i\rangle\langle-\alpha_i|$, $i = 1$,~2.
It is clear from the above inequality that for $\alpha_1=0$ or $\alpha_2=0$ we get identity. It is due to the density operator~(\ref{two:mode}) in this cases has the factorized form.

Let us take the density operator in the form which corresponds to mixture of a thermal and coherent states
\be\label{therm}
\widehat{\rho}\,(1,2)=\frac{1}{2}\,|\alpha_1\rangle \langle\alpha_1|\otimes \widehat{\rho}_{T}(2)+\frac{1}{2}\,\widehat{\rho}_{T}(1)\otimes |\alpha_2\rangle \langle\alpha_2|.
\ee
Here $\widehat{\rho}_{T}(i)$ is the thermal state $\widehat{\rho}_{T}(i) = \frac{1}{Z}\,e^{-\left ( \widehat{a}^+_{i}\,\widehat{a}_{i}+\frac{1}{2} \right )/T}$, where $T$ is a temperature, $Z = \mbox{Tr}\left (e^{-\left ( \widehat{a}^+_{i}\,\widehat{a}_{i}+\frac{1}{2} \right )/T}  \right )$ is a partition function. The operators $\widehat{a}_{i}^+$ and $\widehat{a}_{i}$ are the photon creation and annihilation operators of $i$th mode ($i = 1$,~2), respectively. Taking into account $\mbox{Tr}\,|\alpha_{i}\rangle\langle\alpha_{i}|\,\widehat{\rho}_{T}(i) = \langle\alpha_{i}|\widehat{\rho}_{T}(i)|\alpha_{i}\rangle = \frac{1}{1+N}e^{-\frac{|\alpha_{i}|^2}{1+N}}$,
where $N=(e^{1/T}-1)^{-1}$ is the mean photon number, we obtain the explicit form of the purity inequality
\be\label{prov:therm}
 1+\mu\,(12)-\mu\,(1)-\mu\,(2)=\frac{1}{2}\left ( 1-\frac{1}{1+N}e^{-\frac{|\alpha_1|^2}{1+N}} \right )\left ( 1-\frac{1}{1+N}e^{-\frac{|\alpha_2|^2}{1+N}} \right )\geq 0.
 \ee 
One can see that the inequality is fulfilled for all the parameters determining the state~(\ref{therm}). 

\section{Conclusions}
\pst 
To conclude we formulate the results obtained in this work. We reviewed the properties of even and odd coherent states introduced in~\cite{9} for one dimensional oscillation and~\cite{12,13} for multidimentional oscillator. The replica method and properties of the von Neumann entropy were studied and applied to the mixed even and odd superposition states. The purity characteristics of the even and odd mixed Schr\"{o}dinger cat states were considered. In this paper, we obtained in explicit form the von Neumann entropy for mixture of coherent states. The explicit expression Eq.~(\ref{entrexpl}) for the von Neumann entropy of one-mode field state given in terms of mixture of coherent state is obtained. Analogously the entropy of the reduced density operator $\widehat{\rho}\,(1)$ of two-mode field state Eq.~(\ref{cats}) is obtained and given by Eqs.~(\ref{entr1}),~(\ref{De}). In addition, purity inequality for mixture of two-mode even and odd Schr\"{o}dinger cat states was considered and checked on example of the two- mode cat states (see Eq.~(\ref{ineq:prov})). In future publication the application of the approach will be performed for nonlinear f- oscillator~\cite{24} and  its even- odd Schr\"{o}dinger cat states~\cite{25,26} as well as application in the theory of the amplifier.

\section*{Acknowledgments}
\pst
 The authors wish to thank Prof. S. N. Filippov for some useful discussions.

\end{document}